\documentclass[
    reprint,amsmath,amssymb,aps,floatfix,footinbib
]{revtex4-2}

\usepackage[dvipsnames]{xcolor}
\usepackage[polish,english]{babel}
\usepackage[utf8]{inputenc}
\usepackage[T1]{fontenc}
\usepackage{graphicx}%
\graphicspath{{images/}} %
\usepackage[caption=false]{subfig} %
\usepackage{dcolumn}%
\usepackage{bm}%
\usepackage{hyperref}%
\usepackage{amsmath}
\usepackage{appendix}
\usepackage{amsthm}
\usepackage{amssymb}
\usepackage{multirow}
\usepackage{booktabs}
\usepackage{mathrsfs}
\usepackage{mathtools}
\usepackage{rotating}
\usepackage{setspace}
\usepackage{unicode-math}
\usepackage{tikz}
\usepackage{standalone}

\DeclareMathOperator{\sech}{sech}
\DeclareMathOperator{\arctanh}{arctanh}
\DeclareMathOperator{\arcsinh}{arcsinh}

\usepackage{microtype} 

\begin{document}
\singlespacing

\theoremstyle{remark}
\newtheorem*{problem}{Problem}

\theoremstyle{plain}
\newtheorem{theorem}{Theorem}
\newtheorem{lemma}[theorem]{Lemma}
\newtheorem{corollary}[theorem]{Corollary}

\title{A visual introduction to curved geometry for physicists}

\author{Karol Urba\'nski}
\email{karol.j.urbanski@doctoral.uj.edu.pl}
\affiliation{Szkoła Doktorska Nauk Ścisłych i Przyrodniczych\\ Institute of Physics, Jagiellonian University in Kraków.}

\begin{abstract}
    This article provides a gentle, visual introduction to the basic concepts of differential geometry appropriate for students familiar with special relativity. Visual methods are used to explain basics of differential geometry and build intuition for all types of Riemannian and Lorentzian manifolds of constant curvature. A visual derivation of the Thomas precession is given, showcasing the utility of differential geometry while also pointing a spotlight at certain intricacies of Minkowski space crucial from a pedagogical perspective. In addition, a straightforward method to generate some Carter-Penrose diagrams -- suitable for students with no differential geometry knowledge -- is presented, and a new method of indicating distortion on spacetime diagrams is shown.
\end{abstract}

\maketitle

\section{Introduction and rationale for the article}
General relativity is one of the two pillars on which modern physics rest. As a consequence, it is inevitable for physics students to have to learn the complicated mathematical machinery it needs. Here we encounter a lot of forks in the road. Some references focus on the mathematics -- mainly differential geometry books for mathematicians, though it is also the approach taken by \textit{Spacetime and Geometry} \cite{carroll2019spacetime} and \textit{General Relativity} \cite{wald2010general}. For individuals more physically minded, there are approaches where physics are done first -- one such approach is the evergreen tome \textit{Gravity} \cite{hartle2021gravity}, which uses manipulations of metrics to obtain physical results \textit{before} the underlying mathematics are explained; the resulting tome is a relativity textbook appropriate for undergrads. That is no easy task. And let us not forget about the Black Bible: Misner, Thorne and Wheeler's \textit{Gravitation} \cite{misner1973gravitation}, the greatest challenge to desk construction a relativist can find. A wonderful tome with beautiful prose, focusing on imparting a geometric and visual thinking, in my mind it has only two flaws. The first one is that it's hard to meet people who have read it from cover to cover, probably owing to its intimidating size. A typical relativist only uses it as the Final Reference for a concept they look up in it's index.

The second flaw, however, is much more surprising for such a geometry-first tome. In fact, all the aforementioned references have this problem: in each of them, the business of actually imagining a Riemannian or Lorentzian space, immersing ourselves in it and finding some sort of intuition is sidestepped. The off-ramp taken is typically the language of tensor calculus: even Hartle, despite being aimed at a relativity layperson, reaches for the line element formalism immediately.

This is not surprising, and even hard to argue against. After all, curved spacetime is difficult. This means that we certainly need to learn all those tools, and at a high level of abstraction algebra is our only ally. Nevertheless, what is missing is a unified reference that allows a student to have some \textbf{inkling} of how to imagine spaces of curvature.

To wit, if you're a seasoned physicist, consider this question: how do you imagine a flat surface? What about positively curved? And what about negatively curved? 

In all likelihood, what you've imagined is a plane, then a sphere or a point on a dome, and a saddle. This is the standard way to imagine these concepts, based on the definition of Gaussian curvature. The principal curvatures $\kappa_1$ and $\kappa_2$ along orthogonal axes combine to create Gaussian curvature
\begin{equation}\label{eq:gaussian-curvature}
    K = \kappa_1 \kappa_2,
\end{equation}
and so picturing a positive curvature as a dome and a negative curvature as a saddle (two curves of opposite curvature orientation crossing) is natural. But there's a problem: those visualisations show the vicinity of \textbf{a single point}. Therefore, you cannot try to imagine a larger surface and expect your intuition to work right. In fact, hyperbolic space is not even embeddable in $E^3$ in it's entirety; mathematicians often resort to using various models of hyperbolic space to visualise a portion of it. In general, each of these models requires certain `rules' to be used correctly, and drawing geodesics on them is not trivial. This means they are often not viable for showcase to new students, and discarded the moment tensor calculus is introduced. By the time students are done with their differential geometry course, they often would rather use algebraic methods rather than 'regress' to geometric intuition. However, geometric intuition allows a scientist to reject spurious hypotheses and pick the interesting ones for further research. Even the great geometer David Hilbert thought so \cite{hilbert2021geometry}. Not having those tools is a self-imposed handicap.

This article serves to bridge that gap. It is an introduction to differential geometry for undergraduate physicists. Physicists taking their first general relativity course have one crucial advantage over mathematicians taking their first differential geometry course: they have usually already been taught and understand special relativity. This means they have an intuitive understanding of Minkowski space and Minkowski diagrams. Hyperbolic space is perfectly embeddable in three dimensional Minkowski spacetime. For that matter, so are both Lorentzian surfaces of constant curvature, which are of arguably greater interest to physicists -- and yet somehow seldom visualised at all.

This article's goal is to introduce every of those constant curvature surfaces. We will construct them with elementary geometry, and operate on them with physical insight. With luck, this will allow students to enter a differential geometry course with an intuitive understanding of curved space and spacetime, and endow them with the ability to visualise these geometries when needed. Despite the modest depth of the mathematical toolbox available to a student beginning their general relativity journey, we will also make interesting physics predictions while at it.

\subsection{Notes on using this article}
This article showcases a clear sequence of visualisations and their consequences, assuming some special relativity knowledge prerequisite, and no differential geometry knowledge. In principle, it could be shown as is at the beginning of a differential geometry course. However, I would recommend interspersing these visualisations with more traditional introductions to mathematics of differential geometry. For instance, after spherical geometry, it would be appropriate to talk in more depth about concepts such as Gaussian curvatures. The intuitive explanation for parallel transport can be a starting point for introducing the Levi-Civita connection. After the introduction of the de Sitter space, it would be appropriate to talk about the Riemann curvature tensor and geodesic deviation, et cetera.

If this were to be a comprehensive source of all relevant differential geometry material, it would be a hefty book, not an article. Therefore, I would suggest using it as a repository to draw ideas for use in a larger differential geometry course, rather than a rigid lecture plan. The lecturers should be deliberate about how they integrate this material with their own general relativity/differential geometry curriculum. That being said, it will do fine as an intro for curious students. It is also appropriate for a short seminar for students who have already taken their differential geometry course. They will be well equipped to understand many of the used concepts, and in turn will learn how to imagine these geometric spaces to bolster their algebraic toolbox.

\section{Curved geometry}
Calculations in curved geometry are the backbone of general relativity: it is our reigning theory of gravity. It explains this weakest of the four fundamental forces as a consequence of the curvature of spacetime geometry around massive objects.

To be capable of making predictions in general relativity, we need a good deal of complicated mathematics. The mathematical tools we use -- called differential geometry -- have to be adapted to a background that curves and twists, and responds itself to this curving and twisting. We shall begin our exploration of this concept by analysing the simplest curved two dimensional surface that we are all familiar with: a sphere.

\begin{figure}
    \fbox{\includegraphics[width=0.97\linewidth]{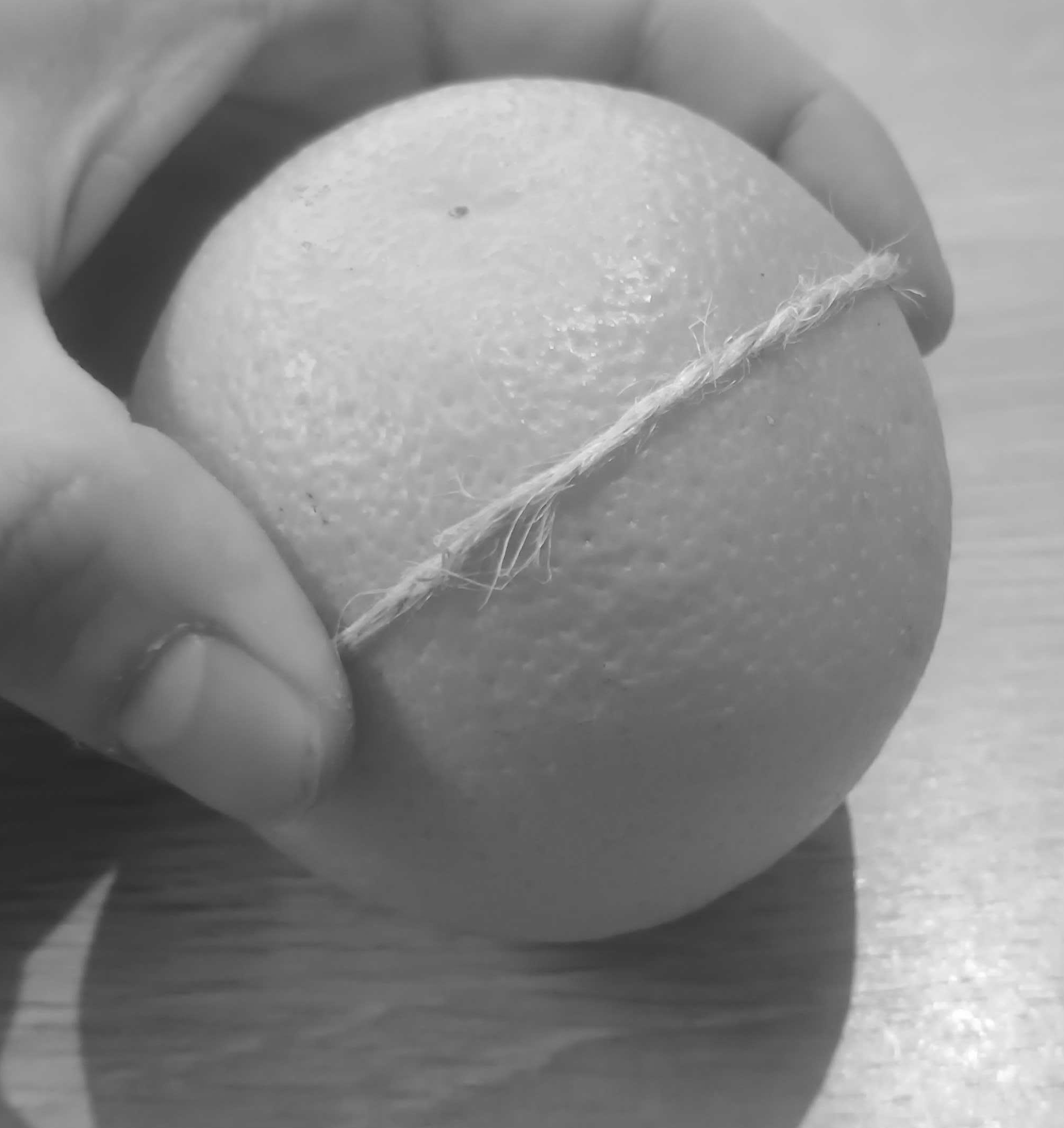}}
    \caption{When pulled taut, a length of string will necessarily pick the shortest distance between two points on a sphere. This is a geodesic.}
    \label{fig:geodesic-string}
\end{figure}

A sphere is described in three dimensional Euclidean space $E^3$ with a parametric equation in $\phi \in [0, 2\pi), \theta \in (-\pi/2, \pi/2)$:
\begin{equation}
    x^i = (\sin{\theta} \cos{\phi}, \sin{\theta} \sin{\phi}, \cos{\theta}).
\end{equation}
Such a surface is quite different from Euclidean geometry. In the beginning of his \textit{Elements} \cite{euclid300bc}, Euclid gives definitions and axioms required to study shapes on paper. One of them is the axiom called the parallel postulate, which can be phrased in modern form as: there is exactly one line that can be drawn parallel to another given one through an external point.

Among the axioms concerning straight lines, circles, and equality of all right angles, this particular one intrigued mathematicians most. Many attempted to find a way to prove it from the other axioms for millennia, to no avail. And for good reason: when we assume something different, we can obtain a different, yet valid geometry.

In particular, on the surface of a sphere, any line is a great circle, and it will intersect with every other great circle: a `straight line' that doesn't cross our line can never be found. A way to discover this in a practical way is to take a spherical object, such as an orange or a globe, ideally with drawn circles of longitude and latitude; then, play around with a bit of string. See figure \ref{fig:geodesic-string} for a showcase of the analog to a straight line in curved spacetime: a geodesic.

\begin{figure}
    \fbox{\includegraphics[width=0.97\linewidth]{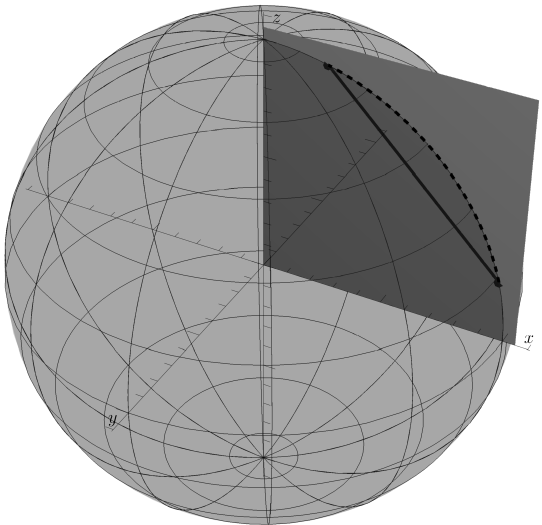}}
    \caption{The construction of a geodesic (dashed line) between two points as the intersection of the plane spanning the origin and the two points with the sphere's surface. The thick line is the closest distance in the $E^3$ space, which also lies on this plane.}
    \label{fig:geodesic-shadow}
\end{figure}

In the particular situation of the sphere -- which, when embedded in three dimensional space has a center of symmetry -- we can construct a geodesic in another, unusual way. Consider our sphere in 3D space. If it were hollow, the line between two points on it's surface that would be the shortest is simply a straight line between points, and a chord of our sphere. So, when we restrict ourselves to the surface, our geodesic should match this chord as closely as possible. (While the statement `match as closely' is vague in mathematical terms, it can be made rigorous; see Appendix A). This can be done by `projecting' the straight line out from the center of the sphere onto the sphere's interior. This `shadow' gives a geodesic on our constant curvature surface! See figure \ref{fig:geodesic-shadow}. More mathematically, we can take the unique plane given by the origin and our two points: it's intersection with the sphere will be our geodesic. While unusual, we will see that this construction will benefit us a great deal for visualisation purposes in later sections: it works with every embedded surface of constant curvature.

\section{Parallel transport -- an intuitive explanation}

Two dimensional beings on the surface of a sphere can, at a point, draw vectors tangent to the sphere. If the curvature is sufficiently small (aka: the sphere sufficiently big in comparison to the inhabitants), the surface in their neighborhood looks flat to the inhabitants. The vector looks as if it were confined to that `flat' plane. \footnote{For humans, the surface of the Earth looks two dimensional, and so when we go on a nature hike, we can use a projection of Earth's surface in our immediate vicinity to a two-dimensional paper page (colloquially known as a `map') to orient ourselves.} This plane at a point $x$ is called, appropriately, the tangent space or the tangent plane $T_x M$. Each point on a curved surface has an associated tangent space. A space (or spacetime) $M$ that looks flat in a sufficiently close neighborhood to each point is called a manifold. When it's tangent space at each point is Euclidean, it is called a Riemannian manifold; when the tangent space is Minkowskian, it is called a Lorentzian manifold.

A question then arises. How does one transport this vector along a curved manifold? If we made no adjustments between different points on a sphere, clearly our vector would start to poke out of the tangent plane at the other point. Therefore, we need to figure out a way to confine it to the surface. The first guess would be to project the vector onto the nearby point's tangent plane; however, that would change our vector's length. The best way to avoid this is to project by means of a rotation. This rotation needs to compensate for the curvature of our surface at the point of transport. For a movement along a geodesic corresponding to the change of angle $d\theta$ on the surface of a sphere, the necessary rotation is obviously a rotation about the vector's origin by $d\theta$. For other curves, it is more complicated, and requires tools of tensor calculus to be performed in the general case. This construction can be looked at in a different way, too: we can envision transforming the tangent plane at the neighbouring point to be squarely `flat' with the original tangent plane, transporting the vector, and then transforming the tangent plane back. This intuition works with non-geodesic curves, also: we need to `flatten' all nearby tangent spaces, but then we can follow a non-geodesic path.

\begin{figure}
    \fbox{\includegraphics[width=0.97\linewidth]{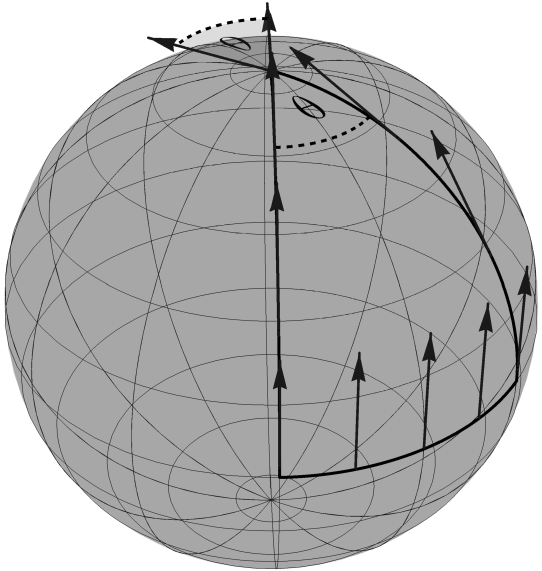}}
    \caption{A particularly simple geodesic path for parallel transport. We see that despite going back to the same place, our vector has rotated $\theta$ degrees, which is exactly the amount of meridians we crossed on the equator.}
    \label{fig:triangle}
\end{figure}

Rather than perform this difficult calculation for an arbitrary curve, let's find a closed path where we can construct the change in an intuitive manner. See figure \ref{fig:triangle}. The construction relies on the fact that the equator is a geodesic and a vector orthogonal to it clearly should remain orthogonal. Following lines of longitude is also quite clear: they are also geodesics, and the angle between the geodesic and our vector should remain constant. This allows us to construct a spherical geometry triangle, which contains two locally right angles and an angle of $\theta$. As we can clearly see in the picture, our vector appears to have rotated by an angle of $\theta$ as well.

A spherical triangle has an excess of angle over the Euclidean sum $\pi$. This angular excess can be used to directly compute a Riemannian surface's curvature using what's called the Gauss-Bonnet theorem:
\begin{equation}\label{eq:gauss-bonnet}
    \sum_i \alpha_i = \pi + \int_T K,
\end{equation}
where $\alpha_i$ are angles of the geodesic triangle, $K$ is the curvature at a point, and $T$ is the interior of the geodesic triangle. For a sphere, the curvature $K$ must clearly be positive. Carrying out this calculation by noting that the triangular path has an excess angle $\theta$ and the interior area $A = \theta * R^2 * 2 \pi = \theta$ we would find that 
\begin{equation}
    \theta = \theta K,
\end{equation}
and so $K = 1$. Looking at the definition \eqref{eq:gaussian-curvature} we can find a connection to the two principal curvatures, which must have the same sign.

With an idea of how to parallel transport, there is a more useful definition of a geodesic; this definition works in a localized manner, which will have implications in more complicated geometries. On a flat Euclidean plane, we can draw a straight line by taking a point and a vector. When anchored at the point, the vector (provided it is non-zero) can construct a new point. We can then (thanks to the flatness of the surface) easily move the vector to the new point, and repeat constructing another point. This way, with sufficiently small steps between points, we construct a straight line. It turns out this definition also works in curved spacetime: a geodesic is constructed by parallel transporting a vector along a direction pointed by itself.

We can make a model that showcases this more vividly, and will allow us to 'construct' geodesics on any curved 2-dimensional surface we encounter in the real world. Let's take a thin strip of sticky tape and attach some toothpicks or matches to it to represent vectors. Clearly, on a flat surface, this reproduces a flat geodesic -- a straight line. But our sticky tape is elastic and thin enough to glue to an arbitrary surface. This way, we can clearly see that a straight line indeed has no choice but become a great circle on a sphere.

\section{Spherical geometry and Foucault's Pendulum}
\begin{figure}
    \fbox{\includegraphics[width=0.97\linewidth]{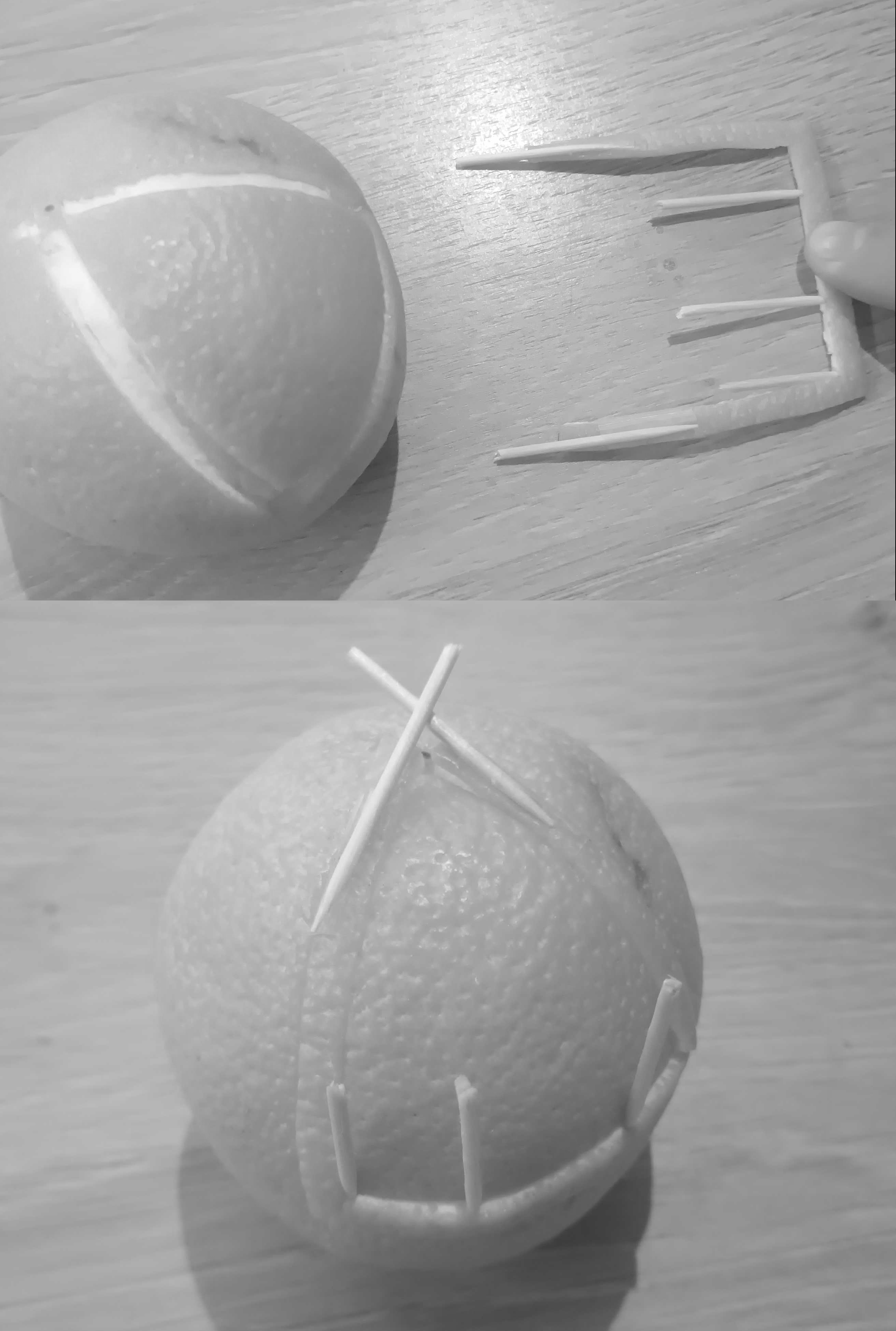}}
    \caption{Top: A strip of orange peel in the shape of a spherical triangle can be flattened on a surface, yielding three straight lines at right angles. Parallel transport around it becomes trivial, and is represented here by toothpicks inserted into the peel. Bottom: When overlaid back onto the surface, we see in spherical geometry our lines make a loop, and the transported vector appears to have rotated.}
    \label{fig:parallel-transport-fruit}
\end{figure}

To find how much a non-geodesic loop we follow deflects a vector, we could construct a curve in flat spacetime that matches the intended path when we lay it out on the curved surface. However, for complicated surfaces, the required guesswork would be difficult. Instead, a better way would be to take the strip of a curved surface and 'flatten it out'. This takes advantage of us envisioning the transport as adjusting the tangent plane and performing the transport in flat space.

Once on a flat plane, it is trivial to attach our toothpick vectors to the curve; all of them parallel to our original vector, since in flat space parallel transport is a simple translation. Then, all we need is to put our curve back onto the surface; see figure \ref{fig:parallel-transport-fruit} for an example that shows this using our original triangle.

\begin{figure}
    \fbox{\includegraphics[width=0.97\linewidth]{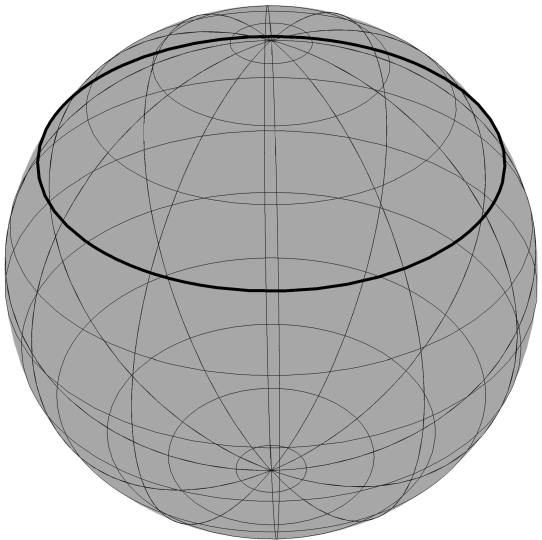}}
    \caption{When placed on a rotating Earth at a certain latitude, the Earth's rotation makes the laboratory perform an apparent movement in the inertial frame. The path is clearly non-geodesic, therefore we should expect a phase shift.}
    \label{fig:foucault}
\end{figure}

We can use this to derive the behavior of Foucault's pendulum. A Foucault's pendulum rotates because when on a circle of latitude that isn't the equator, the rotation of the Earth makes the point in which we perform the experiment travel along a non-geodesic. See figure \ref{fig:foucault} for the diagram of the experiment. Since this curve in general is not a geodesic, our vector (which we identify with the starting direction of the pendulum's movement) will begin to drift from the point of view of our tangent space!

\begin{figure}
    \fbox{\includegraphics[width=0.97\linewidth]{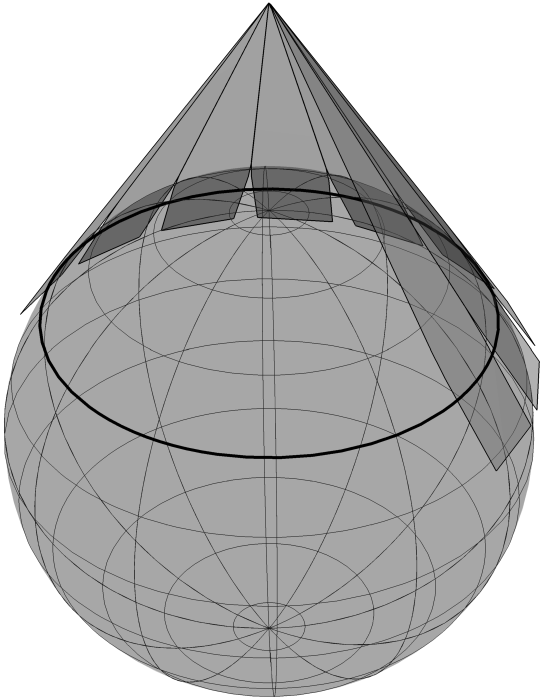}}
    \caption{When individual tangent spaces are plotted along the path, we can extend them upwards to create a cone. A cone is a flat surface that can be easily unrolled onto a flat surface, allowing us to calculate the expected angle directly.}
    \label{fig:foucault-cone}
\end{figure}

Now, we could actually do our little `surgery' and obtain a deficit of angle for an arbitrary curve.\footnote{Curious students are certainly welcome to try performing this `calculation' by hand, by cutting out a thin strip of orange peel and laying it out on a flat surface; make sure to handle any necessary blades with care! The answer can then be directly measured with a protractor.} However, in the case of Foucault's pendulum, there is a degree of symmetry. When we draw the tangent planes in the surface and extend them towards the pole, we will see that they intersect, forming a cone, as in figure \ref{fig:foucault-cone}.

A cone is actually a flat surface with a 'defect'. Crucially for us, we can easily imagine unfolding it onto a surface. When unfolded, we immediately see how much our vector will turn. All that is required is to find what this angle will be depending on the latitude we perform our experiment on, which turns out to be an elementary trigonometric calculation. The angle of latitude $\theta$ is clearly also the cone angle between the height and the slant height. For a cone of slant height $R$ and base radius $r$, once unfolded the angle $\omega$ will be
\begin{equation}
    \omega = 2 \pi \frac{l_{\textrm{base}}}{l_{\textrm{circle}}},
\end{equation}
where $l_{\textrm{circle}} = 2 \pi R$ is the circumference of a circle with radius equal to the slant height $R$, ie. we find what proportion of this circle our base of the cone takes.

The circumference of the base is (using the right triangle between the base, height and slant height):
\begin{equation}
    l_{\textrm{base}} = 2 \pi r = 2 \pi \sin{\theta} R,
\end{equation}
and so substituting quantities we obtain the precession of the tangent space, and hence the amount our vector will deflect in one sidereal day on latitude $\theta$
\begin{equation}
    \omega = 2 \pi sin(\theta),
\end{equation}
which is the direct description of Foucault's pendulum.

\section{Minkowski 1+2 phase space diagram}
Having introduced curved surfaces of positive curvature with the example of a sphere, let's turn our attention to envisioning a negative curvature surface. However, it turns out embedding a surface of negative curvature is impossible in three dimensional Euclidean space. But, we have a trick up our sleeve: Minkowski spacetime provides a model of flat space that can be drawn, but allows a negative curvature surface to be embedded.

Let's recall a Minkowski diagram. Actually, since we need a three-dimensional space to immerse a two dimensional surface, we will make use of a 1+2 Minkowski diagram. Further, we will use a phase diagram, at least for representing hyperbolic space. A phase diagram is obtained by relabelling of the $t$ axis to $E = mt$ and the spatial axes to $p_x = m x$, $p_y = m y$. We can easily represent a four-momentum vector of timelike, null and spacelike vectors on such a diagram. We will also make use of the concept of rapidity: Minkowski spacetime is well described using symmetries of hyperbolic rotations \cite{urbanski2025visual}. On these phase diagrams, we can find very simple trigonometric relations for a hyperbolic triangle with a hyperbolic angle (called rapidity) $\zeta$:

\begin{align}
    \beta &= \tanh{\zeta}, \\
    E &= \gamma m = \cosh{\zeta} m, \\
    |p| &= \beta \gamma m = \sinh{\zeta} m.
\end{align}

\begin{figure}[b]
    \fbox{\includegraphics[width=0.97\linewidth]{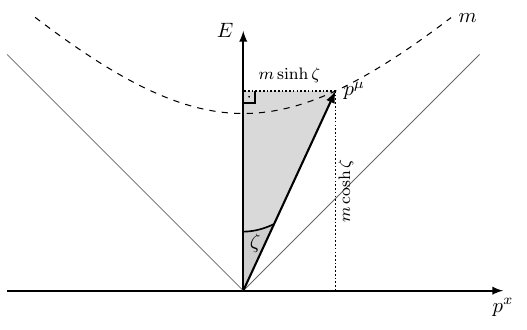}}
\caption{$p^\mu$ is a four-momentum vector of a particle of mass $m$. It starts at the origin, and ends on a hyperbola of hyperbolic radius $m$. Lorentz boosts of the frame may move the momentum vector, but it must remain on the shell. Marked is the angle $\zeta$ corresponding to the rapidity of the particle. Marked with dotted lines are the values of four-momentum components $E = m \cosh \zeta$ and $p = m \sinh \zeta$, and the light cone. The shaded triangle is an example of a right hyperbolic triangle subtended by the four-momentum vector, the $E$ component and the $p$ component.}
    \label{fig:minkowski-triangle}
\end{figure}

See figure \ref{fig:minkowski-triangle} for an example 1+1 Minkowski right hyperbolic triangle and it's properties.

\section{Hyperbolic geometry as consequence of velocity addition: construction of a hyperboloid}
Let's give some special attention to a certain surface we can find on a phase diagram: the particle's mass shell. Clearly, in the particle's own frame of reference, it will occupy the spot $(m, 0, 0)$ on our diagram. In a different frame of reference, in accordance with special relativity, for a rapidity $\zeta$ and direction of boost $\phi$ we get
\begin{equation}\label{eq:mass-sheet}
    p^\mu = (m \cosh{\zeta}, m \sinh{\zeta} \cos{\phi}, m \sinh{\zeta} \sin{\phi}).
\end{equation}

\begin{figure}
    \fbox{\includegraphics[width=0.97\linewidth]{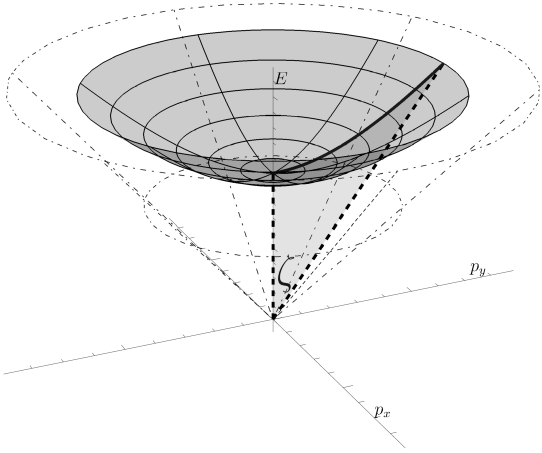}}
    \caption{Constructing rapidity space by tracing out the mass shell with increasing rapidity $\zeta$ and then rotating it.}
    \label{fig:hyperboloid}
\end{figure}

This means the end of a vector of a particle of mass $m$ must occupy a point on the upper sheet of the hyperboloid described by \eqref{eq:mass-sheet}, called the \textbf{mass shell}. By construction, it is a surface that curves in a hyperbolic manner away from the center. By the properties of Lorentz transformations points in one angular direction are indistinguishable, and by the properties of spatial rotations points are indistinguishable by their angle. This means it's a maximally symmetric surface. Looking at figure \ref{fig:hyperboloid}, we can also immediately see that it's a spacelike surface, since it can never be angled so as to touch or cross the light cone, it can only asymptotically approach it.

\begin{figure}
    \fbox{\includegraphics[width=0.97\linewidth]{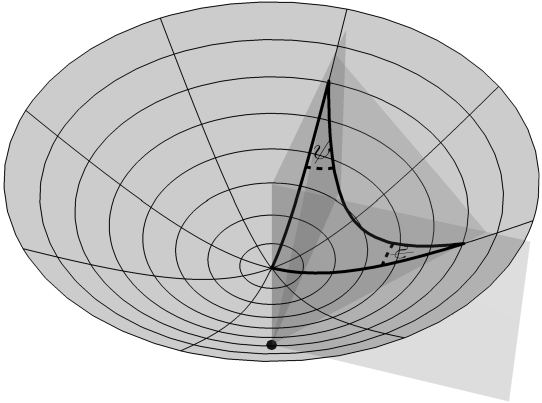}}
    \caption{Constructing a right triangle along the two orthogonal axes and connecting them with a geodesic in a symmetric way creates a triangle in which the angle $\psi$ is obviously smaller than $\pi/4$, therefore the entire triangle's angles combine to less than $\pi/2$.}
    \label{fig:hyperboloid-deficit}
\end{figure}

Finally, because there is a center of Lorentz symmetry at $(0, 0, 0)$, we can repeat a `triangular' construction similar to the one with geodesics. See figure \ref{fig:hyperboloid-deficit}. This time, we construct the geodesics using our plane intersection method. Two of them we construct to lay on lines of constant $\phi$ separated by a right angle $\delta \phi = \pi/2$, and the third one connects two points equidistant from the center. The angles near vertices away from the origin are clearly symmetric, and they clearly are both \textbf{smaller} than $\pi/4$. This means our triangle has an angular deficit, confirming by the Gauss-Bonnet theorem we are on a surface of constant negative curvature.

In Minkowski 1+2 spacetime, this surface of constant curvature is perfectly embeddable! We have obtained it using hyperbolic trigonometry, which is measuring ratios of distances in a space endowed with symmetries of hyperbolic character; this is why this space is often called the hyperbolic space, and this particular model of it the hyperboloid model. 

There exist other models -- you may be familiar with the pseudosphere, the Beltrami half-plane or the Poincare disc. But for a relativist, the hyperboloid model is the natural choice, because it behaves in a simple manner under Lorentz transformations: a Lorentz transformation maps a hyperboloid onto itself.

\begin{figure}
    \fbox{\includegraphics[width=0.97\linewidth]{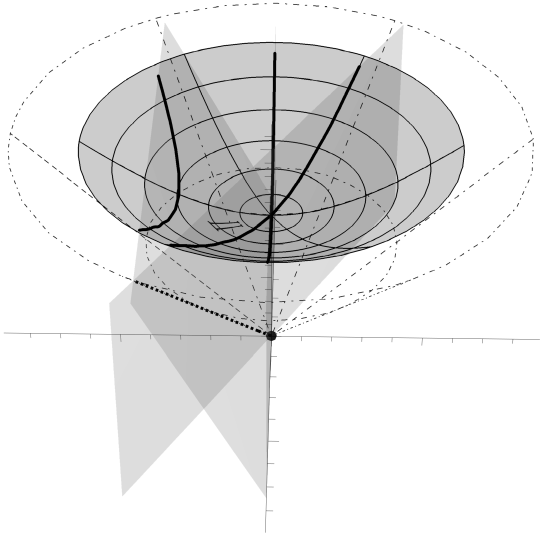}}
    \caption{It's possible to construct multiple geodesics through a point $p$ that are parallel to a geodesic not through that point. The dashed line showcases the intersection of generative planes corresponding to the limiting asymptotically parallel geodesic.}
    \label{fig:hyperboloid-angle}
\end{figure}

It also allows for drawing geodesics simply, which can be very handy. Consider figure \ref{fig:hyperboloid-angle}. In it, we construct a geodesic that doesn't cross the center of the hyperboloid, offset by some rapidity $\zeta$. Then, we can see quite easily that in addition to the obviously parallel geodesic drawn with the plane through the center, we can find geodesics that won't cross the original geodesic in a range of angles: all we need to do is draw our plane so that it intersects the plane of the original geodesic outside the light cone. Then, the geodesics can never cross. The figure shows the limiting parallel geodesic, in which the planes intersect exactly along the light cone. We can use this to find this maximum angle.

The angle $\Pi$, called the angle of parallelism, is the highest angle for a triangle consisting of a geodesic, a perpendicular side of distance $\zeta$ and the asymptotically parallel side. From the positioning of a chosen arbitrary point where the plane constructing our initial geodesic intersects the light cone, we can see the angle's cosine will be the point's spatial coordinate in the Minkowski space of immersion divided by the point's time coordinate:
\begin{equation}
    \cos{\Pi} = \frac{\sinh{\zeta}}{\cosh{\zeta}} = \tanh{\zeta}.
\end{equation}

This showcases that we can draw an infinite family of geodesics parallel to another geodesic through a point, which is the hyperbolic geometry's parallel postulate counterpart. This method of finding the points of intersection of geodesics is also adaptable to arbitrary geodesics: the generating planes give a line of intersection through 1+2 Minkowski space; the intersection of that line with the hyperboloid will by construction give the point of geodesic collision.

\subsection{Derivation of Thomas precession}
Our surface showcases that velocity addition in relativity is complicated: by construction, it must be the surface describing the `rapidity' space. It's $E$ coordinate is actually $m \gamma$ at each point, showcasing that the Lorentz factor is simply a proxy for rapidity. And rapidity, by the relation (easily verifiable with hyperbolic trigonometry)
\begin{equation}\label{eq:tanhzeta}
    \tanh{\zeta} = \beta = \frac{v}{c}
\end{equation}
is one description of relativistic velocity. This helps contextualize why velocity addition looks different than in Newtonian physics: we are bound by the non-zero curvature of rapidity space. The angular deficit showcased actually has a manifestation as a rotation when non-collinear Lorentz boosts are performed, as seen in a previous AJP article \cite{aravind1997wigner}.

So what would happen if, through non-inertial motion, we would be forced to draw a smooth non-geodesic curve through rapidity space?

\begin{figure}
    \fbox{\includegraphics[width=0.97\linewidth]{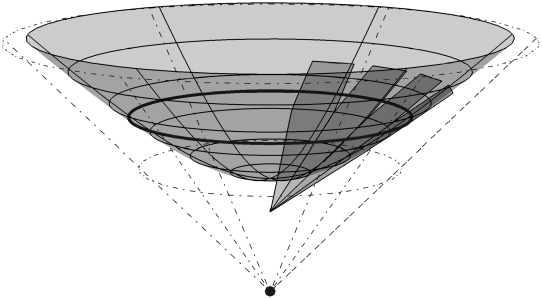}}
    \caption{A circular orbit at relativistic velocities will look like a circle on rapidity space, with radius $\zeta$ corresponding to the movement's rapidity. We can draw tangent spaces at each point of the orbit through rapidity space. Their intersection is a spacelike conical surface.}
    \label{fig:hyperboloid-cone}
\end{figure}

The answer is that we would also expect a phase shift, exactly like in the case of the Foucault pendulum. This effect is called Thomas precession. Let's pick a very symmetrical curve through the phase space, associated with circular motion with a constant rapidity $\zeta$. At the start of our motion, starting in the direction $x$, $p^\mu = m( \cosh{\zeta}, m \sinh{\zeta}, 0)$. Through smooth turning, we will trace out a circle in rapidity space until we get back to the starting point the moment we have travelled the entire circumference of our path. This curve is clearly non-geodesic in rapidity space, and we can once again draw tangent spaces at each point and construct a cone; see figure \ref{fig:hyperboloid-cone}.

To obtain the unfolding angle $\delta \psi$, we take the cosine of the angle $\alpha$ between the height and the slant height. This can be obtained by remembering that the tangent of the angle $\alpha$ of a tangent line to a function is it's slope -- it's derivative at a point corresponding to $x = \sinh{\zeta}$:
\begin{equation}
    \tan{\alpha} = (\sqrt{1 + x^2})' = \tanh{\zeta} = \beta,
\end{equation}
so in the end we get
\begin{equation}\label{eq:wrong-thomas}
    \delta \psi = 2\pi \cos{(\arctan{\beta})} = \frac{1}{\sqrt{1 + \beta^2}}.
\end{equation}

This is... not quite right.  What we would expect as an answer for the angle is $\delta \psi = 2\pi \gamma$. Our expression \eqref{eq:wrong-thomas} is close, but no cigar. So what happened? Before we answer that, let's notice something interesting: if we substitute $\beta \rightarrow i \beta$, we get the expected answer!

\subsection{$i dt$ -- the (imaginary) elephant in a Minkowski room}
It turns out that whenever we take angles or do standard trigonometry on a chart like we've just done, what we're functionally doing is insert an extra hidden step. After drawing all our surfaces and tangent spaces and cones in Minkowski, by using regular trigonometry to measure distances in the vertical directions, we've then embedded that Minkowski in Euclidean space. When done deliberately, it is often called a Wick rotation. But we have to be careful, because it is also the dreaded $-dt'^2 = (i dt)^2$ trick:

\begin{equation}
    \beta = \frac{v}{c} = \frac{x}{i c t} = - i \frac{v_{\textrm{euclidean}}}{c}
\end{equation}

There are very good reasons not to do this, many enumerated in \cite{misner1973gravitation}. The most relevant to us is that it hides the hyperbolic structure of Minkowski space, and fools us into thinking we can operate on Minkowski spacetime as if it were Euclidean space. But, as showcased here, when we refer to an 'angle' on a diagram on a Lorentzian plane through Minkowski -- by using ordinary trigonometric functions directly -- we are actually doing this. In fact, when searching for meaning of angles on Minkowski diagrams either in Google, Wikipedia or an LLM, the primary answer given is often the explanation
\begin{equation}
    \tan{\theta} = \beta = \frac{v}{c} = \frac{x}{tc},
\end{equation}
but this is mistaken in it's assumptions. We can get away with it in this instance because the tangent function is antisymmetric and this means 'hiding' the imaginary unit gives the correct answer when defining an angle between the $t$ axis and the line/vector of a Minkowski diagram. But this is coincidental\footnote{Or, to be more specific, it is not a mathematical coincidence -- it is a consequence of the properties of the analytic extension of regular trigonometric functions along with the properties of the Minkowski embedding. However, the reasoning we are using to find it is unaware of this, and therefore spurious. This makes it coincidental from a physics and pedagogical standpoint.}. 

Little wonder students often stubbornly attempt comparing lengths on a Minkowski diagram: after all, comparing length ratios is how trigonometry is defined. \footnote{The word comes from Ancient Greek, where it means $\tau\rho\iota\gamma\omega\nu\omicron\nu$ `triangle' and $\mu\epsilon\tau\rho\omicron\nu$ `measure'.} Whenever we use these tricks without explanation, we therefore reinforce the appearance that length ratios on a Minkowski diagram can be compared. Perhaps this is an argument to retire regular trigonometry from special relativistic explanations entirely.

\subsection{Resolution using hyperbolic trigonometry}
Aware of what we're doing, we can use $i$ to obtain the expression for Thomas precession. But, since we don't want to hide the hyperbolic character of Minkowski space, what can we do? The answer is simple: use hyperbolic trigonometry when we're computing ratios of lengths in triangles described by a hyperbolic angle. This means using hyperbolic angles between two vectors of the same type -- either two timelike, or two spacelike lengths in a triangle where the third vector is of the opposite type. That angle has a physical interpretation as relativistic velocity, as shown in equation \eqref{eq:tanhzeta}. \cite{urbanski2025visual}

Some care has to be taken to make sure identities still hold, but in the case of our Thomas precession derivation, it suffices to replace the functions in \eqref{eq:wrong-thomas} with their hyperbolic counterparts. The derivative in this space actually gives the hyperbolic tangent angle as the slope, and the cosine can be replaced with the hyperbolic cosine. Then, using \eqref{eq:tanhzeta}, we get
\begin{equation}
    \delta \psi = 2\pi \cosh{(\arctanh{\beta})} = 2\pi \cosh{\zeta} = 2 \pi \gamma,
\end{equation}
yielding the correct answer for the total amount of rotation instantly. Thomas precession -- in the sense of phase drift -- is this angle with $2 \pi$ taken away:

\begin{equation}
    \delta \psi_T = 2 \pi (\gamma - 1).
\end{equation}

\section{Lorentzian surfaces}
\subsection{Two dimensional de Sitter space and its Carter-Penrose diagram}
\begin{figure}
    \fbox{\includegraphics[width=0.97\linewidth]{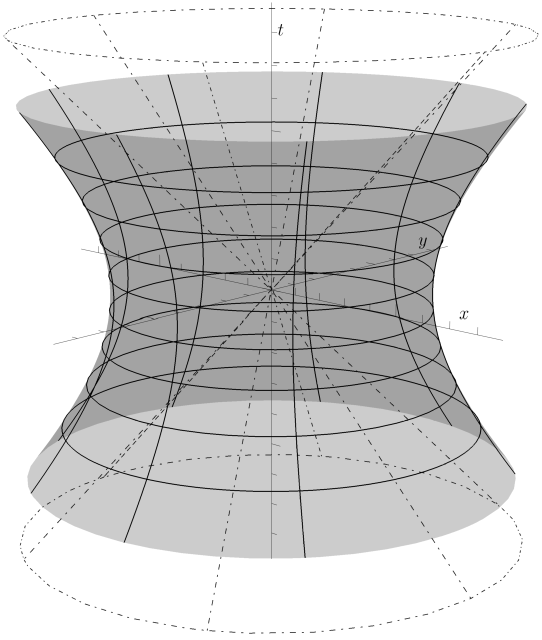}}
    \caption{The timelike surface of a hyperboloid in Minkowski 1+2 is the de Sitter space.}
    \label{fig:ds2}
\end{figure}

The simplest of our pseudo-Riemannian surfaces of constant scalar curvature will be the spacetime counterpart to the hyperboloid. When we rotate hyperbolas to create a surface of revolution, we actually have three choices for the hyperbola we rotate. We've seen the one in the future timelike direction from the origin: it is hyperbolic space. The one in the past timelike direction is exactly isomorphic to this space as well. But we do have a third choice -- the hyperbola(s) in the spacelike quadrants of Minkowski 1+1 become a tubular, hyperbolic sheet, as in figure \ref{fig:ds2}. It's parametric equation is
\begin{equation}
    x^\mu = (\cosh{\zeta} \cos{\theta}, \cosh{\zeta} \sin{\theta}, \sinh{\zeta}).
\end{equation}
By construction, this sheet is Lorentzian (as can be easily seen on the equator, where the tangent space is an ordinary Minkowski plane) and has to have constant curvature. This is a 2 dimensional de Sitter space $dS_2$.

\begin{figure}
    \fbox{\includegraphics[width=0.97\linewidth]{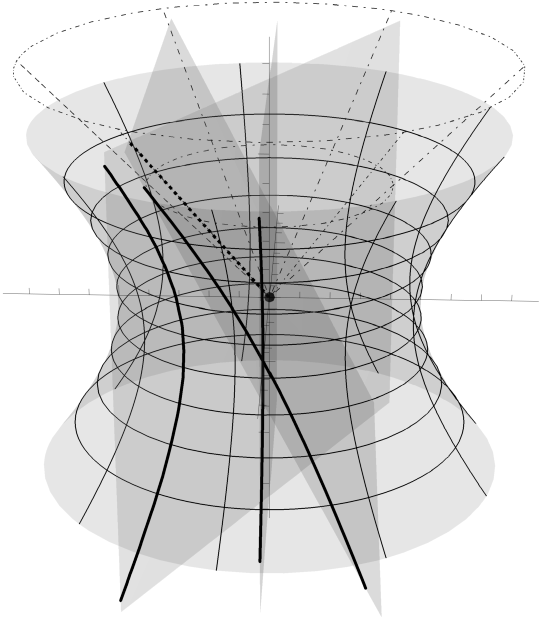}}
    \caption{Just like in the case of the hyperboloid, we can show that there exist multiple parallel timelike worldlines. This indicates that the space 'expands', and explains how the de Sitter spacetime can be a valid model for the future of the expanding Universe.}
    \label{fig:ds2-geodesics}
\end{figure}

The figure \ref{fig:ds2-geodesics} shows that we can find more than one `parallel' worldline through an event in $dS_2$ space to any timelike worldline. Much like with the hyperboloid, we could find the hyperbolic angle $\Zeta$ that corresponds to the asymptotically parallel line. Once again, we have to look at the limiting intersection of the geodesic's constructing plane with the light cone. Since this intersection of planes mirrors the hyperboloid problem, we find
\begin{equation}
    \tanh{\Zeta} = \cos{(\frac{\pi}{2} - \phi)} = \sin{\phi}.
\end{equation}
This hyperbolic angle has the following physical interpretation: two distant observers that are `resting' with respect to the metric coordinates (called fiducial observers) will see each other recede. If one of them desired to `catch' the other one, this hyperbolic angle $\Zeta$ is the rapidity of the minimal boost the observer needs to perform to eventually catch up if the closest distance they ever approach is $\phi$. However, while it seems like this interpretation is simple, it can be very tricky to define a consistent measure of spacelike distance in spaces where observers diverge in this manner. For this reason, it is not as commonly defined as the angle of parallelism in hyperbolic geometry.

\begin{figure}
    \fbox{\includegraphics[width=0.97\linewidth]{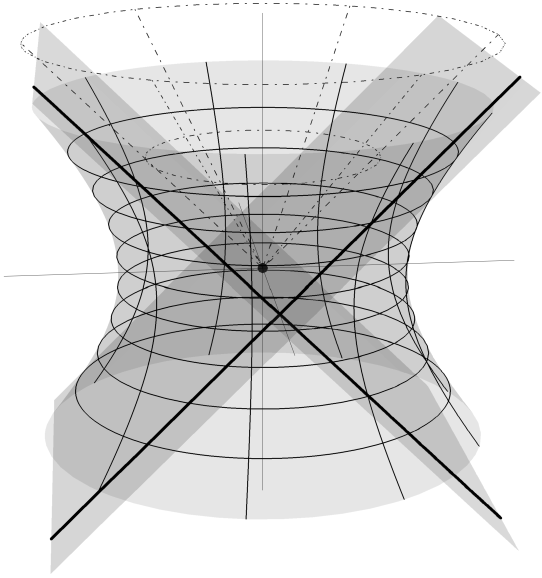}}
    \caption{We can easily construct null geodesics as the intersection with a null plane. The resulting geodesic is a straight line in Minkowski 1+2 space; from the point of view of the $\phi$ and $\zeta$ coordinates on the surface it approaches, but never reaches $\phi = \pm \frac{\pi}{2}$.}
    \label{fig:ds2cone}
\end{figure}

One question that springs to mind is, can we figure out a phase effect similar to what we've done in Riemannian geometries of constant curvature? Sadly, there is no meaningful way to carry a timelike observer on a circle, which limits our ability to find a physically meaningful problem to calculate with our 'cone to find phase drift' tricks. But we can employ our unusual way for finding geodesics, and intentionally construct \textbf{null} geodesics. This is very simple: in the immediate vicinity of the equator, we can find a point in tangent space that corresponds locally to a null geodesic on that tangent space from a starting point. Essentially, we draw a light cone in that tangent space. The origin, point on the equator and point on the tangent space give us a plane that's perfectly angled between the $t$ and $x$ coordinate; a null plane in 1+2 Minkowski space. It's intersection with the de Sitter hyperboloid gives us a geodesic, which must be null by the symmetries of the embedded surface inherited from the embedding space. We can then repeat this construction with the second null geodesic from that initial point and obtain a light cone in de Sitter space; see figure \ref{fig:ds2cone}.

Now, when we look closely, we've actually defined some structure of dS space with respect to the point we've chosen (indistinguishable by symmetry from other points on this manifold). The 'west' patch will never get information from the 'east' patch and vice versa; observers in the 'north' patch can receive signals from both those patches while observers in the 'south' patch can send signals to both. East and west do connect as if on a torus, but they are too far away for information to pass between them the other way. And there is even a patch of spacetime that will never be in contact with our observer. Seasoned relativists will note this slicing is very similar to a Carter-Penrose diagram, and in fact we can take this further.

\begin{figure}
    \fbox{\includegraphics[width=0.97\linewidth]{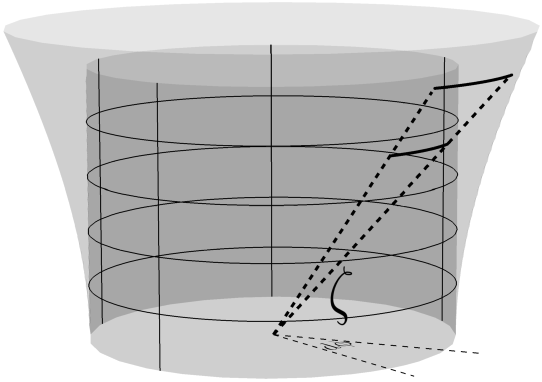}}
    \caption{When a horizontal segment of angle length $d\phi$ at hyperbolic angle $\zeta$ has to be projected onto a unit Minkowskian cylinder, the ratio of the line on the hyperboloid and the line on the cylinder is the horizontal scale factor. Clearly, the scaling will be as the scaling of the entire thick dashed line is to it's length to the cylinder. This ratio is the hyperbolic secant: $k = 1/\cosh{\zeta} = \sech{\zeta}$.}
    \label{fig:mercator}
\end{figure}

With a sphere, there exist ways to project a map that preserves angles onto a cylindrical surface, and then unroll that surface onto a flat plane (since a cylinder is flat):
\begin{align}\label{eq:mercator}
    x(\phi) & = \phi, & x \in (-\pi, \pi), \\
    y(\theta) &= \arctanh{(\sin{\theta})}, & y \in (-\infty, \infty).
\end{align}
Such a map is called a conformal projection map, and with maps of the Earth, it is known as the Mercator projection. It is a specific mapping to a cylindrical surface tangent to the sphere. We can use a hyperbolic Mercator projection analogue for de Sitter space: simply make a Minkowskian cylinder tangent to the 'tube' we have obtained and project onto it conformally. See figure \ref{fig:mercator} for the explanation.

\begin{figure}
    \fbox{\includegraphics[width=0.97\linewidth]{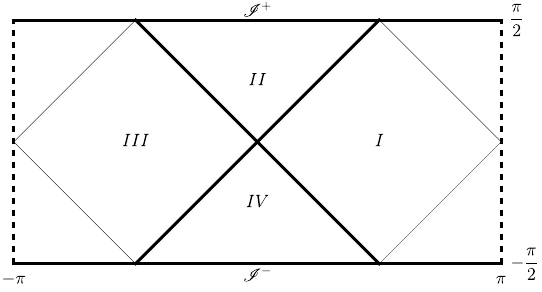}}
    \caption{The Carter-Penrose diagram for $dS_2$ space. The dashed line indicates that the space is joined at these points. Regions I and III can never exchange information between each other. $\mathscr{I}^+$ indicates future null infinity: all null geodesics must end there. Similarly, $\mathscr{I}^-$ is the past null infinity.}
    \label{fig:ds2-penrose}
\end{figure}

After identification of what each surfaces at the 'edge' of the hyperbolic Mercator projection mean -- done using our ability to construct geodesics in figure \ref{fig:ds2-geodesics} -- we obtain the Carter-Penrose diagram (figure \ref{fig:ds2-penrose}) exactly! And we've done it without needing any calculations. What happened? Well, we've built null geodesics -- this is equivalent to taking a null coordinate system on our surface, which is the first step of drawing conformal diagrams. Then, we've transformed it onto a flat surface using an explicitly conformal way; this is the second typical step in drawing conformal diagrams. Essentially, we have repeated the standard derivation, but with a compass and straightedge (adapted to the harsher environment of Minkowski spacetime) instead of algebra.

One important note is that the conformal Mercator projection has a non-trivial algebraic definition on a sphere, and it's no different for the hyperbolic case. That doesn't stop us from a qualitative diagram: we know after all the steps we've taken, the light cone should look like a Minkowski light cone in our final product. This means we can take our null geodesics and know exactly what their shape will be on paper (diagonal lines), without needing to get into minutia. 

That said, the calculations are elementary. The conformality condition implies that vertical and horizontal scaling around each point's vicinity should be the same. The horizontal scaling $k$ from dS onto a Minkowskian cylinder in Minkowski 1+2 space is the ratio of the distances to the cylinder and to the sheet. This is the geometric definition of the hyperbolic secant for the hyperbolic angle
\begin{equation}
    k = \sech{\zeta},
\end{equation}
therefore we set the vertical scaling $h$ to also be
\begin{equation}\label{eq:vert-scaling}
    h = \sech{\zeta}.
\end{equation}
A movement through a small angle $d\phi$ on a parallel on the hyberboloid tube with scalar curvature $1$ corresponds to a change of $k \cosh{\zeta} d\phi$, while a movement of $d\zeta$ along a meridian results in a change of $h d\zeta$. Putting it together we obtain a differential system for our projection
\begin{equation}
    dx = d\phi, \quad dy = \sech{\zeta} d\zeta,
\end{equation}
which we can easily integrate
\begin{equation}
    x(\phi) = \int^\phi_0 d\phi, \quad y(\zeta) = \int^\zeta_0 \sech{\zeta} d\zeta.
\end{equation}
This gives 
\begin{equation}
    x(\phi) = \phi, \quad y(\zeta) = \arctan{(\sinh{\zeta})},
\end{equation}
which, when we remember $\sinh{\zeta}$ in our Minkowski immersion is simply the $t$ coordinate, becomes
\begin{align}\label{eq:hyp-mercator}
    x(\phi) & = \phi, & x \in (-\pi, \pi), \\
    y(\zeta) &= \arctan{(t(\zeta))}, & y \in (-\frac{\pi}{2}, \frac{\pi}{2}).
\end{align}
The expression for the hyperbolic Mercator projection for $dS_2$ is an exact analogue of the spherical Mercator projection expression. We obtain finite limits of angles, allowing the Carter-Penrose diagram to be graphed on a finite strip of paper.

\begin{figure}
    \fbox{\includegraphics[width=0.97\linewidth]{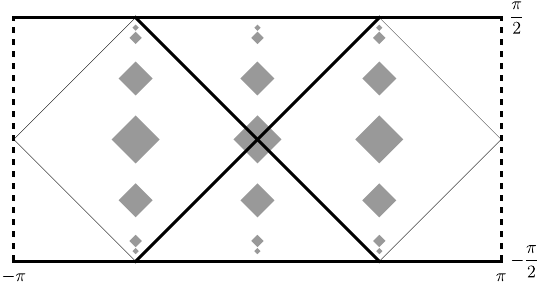}}
    \caption{The conformal factor of a point indicated by a Lorentzian equivalent of the Tissot indicatrix. A causal diamond normalized to correspond to a diamond connecting timelike separated points of some spacetime interval $dl$ appears to have a different size depending on how close it is to the equator, showing the distortion induced by the conformal transformation. The distortion does not impact the angle of the null cone.}
    \label{fig:ds2-penrose-indicatrix}
\end{figure}

On a sphere, the Mercator projection makes things far away from the equatorial plane look bigger. This is because the projected surface is inside the tangent cylinder, and size ratios grow smaller and smaller as we go up on the cylinder. Theorema Egregium is merciless: no mapping of a surface of constant curvature can preserve both angles and areas. Looking at the hyperbolic Mercator projection, we see the same effect but in reverse: our surface is outside the Minkowskian tangent cylinder. So, to showcase this, let's take areas of infinitesimal null diamonds -- which in flat spacetime have an interpretation as the square of the interval, as showcased in previous AJP articles \cite{mermin1998space} and \cite{moriconi2026invariance}. The diamonds on our projected surface grow larger and larger in comparison to a fixed area on the cylinder as we depart the $t = 0$ plane. An infinitesimal causal diamond at a point is a Carter-Penrose diagram equivalent to a Tissot indicatrix, where the length of sides scales with parameter $h$, as shown on the diagram in figure \ref{fig:ds2-penrose-indicatrix}. Using \eqref{eq:hyp-mercator} together with \eqref{eq:vert-scaling}, the conformal scale factor for a point at height $y$ is

\begin{equation}
    \lambda = \sech{(\arcsinh{(\tan{y})})}.
\end{equation}

\subsection{Two dimensional anti-de Sitter space as the $dS_2$ space's ``evil twin''}
\begin{figure}
    \fbox{\includegraphics[width=0.97\linewidth]{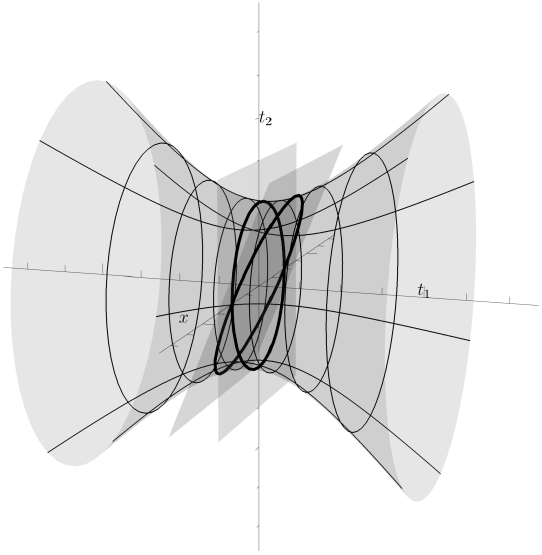}}
    \caption{Anti-de Sitter space embedded into 2+1 Minkowski. Showcased are geodesics from a central point, which have no choice but to intersect on the opposite `poles'. All timelike geodesics tend to approach each other: space appears to contract.}
    \label{fig:ads2}
\end{figure}
We are left with exactly one surface of constant curvature we don't yet know: the anti-de Sitter space. Now, this space doesn't admit embedding into 1+2 Minkowski, but it does admit an embedding into 2+1 Minkowski, as in figure \ref{fig:ads2}. Having these two time dimensions is the only way for the surface to remain Lorentzian everywhere.

\begin{figure}
    \fbox{\includegraphics[width=0.97\linewidth]{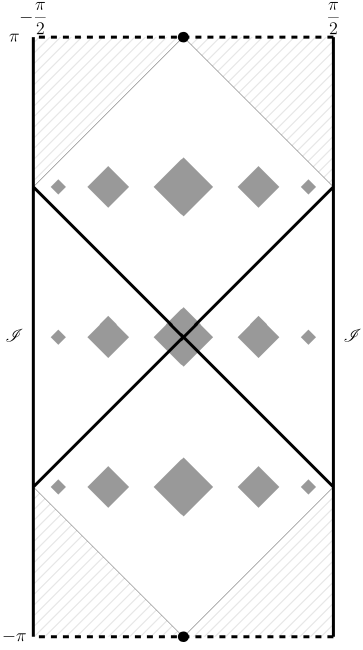}}
    \caption{The Carter-Penrose diagram for $AdS_2$ space. The dashed line indicates that the space is joined at these points. The hatched area indicates an area geodesics starting at the central point can never reach; though it is reachable by other means. The marked point is the intersection point for all timelike geodesics starting at the center. $\mathscr{I}$ are infinities for spacelike geodesics.}
    \label{fig:ads2-penrose}
\end{figure}

As we can see, it looks very similar to $dS_2$ space. In fact, we can also take it further and notice that it's just de Sitter space, but spacelike and timelike directions have been flipped! $AdS_2$ is $dS_2$'s ``evil twin''. Noticing this, we can immediately guess what the Carter-Penrose diagram looks like -- this time we have to project on a cylinder that is oblique, and note the different behavior of timelike geodesics from the central point. The result is shown in figure \ref{fig:ads2-penrose}.

This also means that in $dS_2$ space, if we analysed spacelike geodesics, we would notice a similar effect. This means that whether curvature makes geodesics converge or diverge depends not only on the sign of scalar curvature, but also the character of the geodesics we are researching. This is a fact of great importance in general relativity, where we operate on 4-dimensional Lorentzian manifolds.

In $AdS_2$ space, closed timelike curves exist. This is quite problematic from the point of view of physical theories, since a particle's future can impact it's past. For this reason, in theories where $AdS_2$ spacetime is used -- for instance, AdS/CFT correspondence -- either specific patches of the space are chosen, or a universal covering space is used, so that the diagram is a strip that conntinues infinitely rather than looping onto itself.

\section*{Acknowledgements}
The toothpick idea is inspired by chapter 1 in Tristan Needham's \textit{Visual differential geometry and forms} \cite{needham2021vdgaf}. It is reproduced here so that this article can be a self-contained introduction for use in physics education.

The derivation of Foucault's pendulum can be found in many sources; I have first seen it in \cite{delplace2020geometry}.

I discovered the Thomas precession derivation independently, but in the process of reviewing literature for this article I have found that it has been derived previously in the article in the Russian journal \textit{Physics-Uspekhi} by Krivoruchenko: \cite{krivoruchenko2009rotation}. The content of my article is geared towards a pedagogical approach for a comprehensive introduction of curved geometry visualisations with the geometric phase problem as an illustration. Krivoruchenko's article article presents a rigorous direct approach to the geometric phase problem. Therefore, I recommend it to the curious reader hungry for rigor as the perfect complement to this article.

\begin{appendices}
\section{Why projecting from the center of an immersion works for surfaces of constant curvatures}
One approach is to simply show the congruence of the geodesic equation with the surfaces created by our plane construction.

However, in the geometric vein of this article, let's employ a symmetry argument. The sphere embedded in $E^3$ is invariant under the action of rotations about the center of the sphere. For arbitrary starting point $p$ we can always position it on the pole, and then tracing out geodesics in every direction we construct every possible meridian, and geodesics are orthogonal to circles of constant latitude. (This is Gauss's Lemma in the most trivial possible situation, where it obviously holds). This means that the meridian that intersects the ending point $q$ is the geodesic between these points; but a meridian can also be constructed by an intersection of the plane spanning the origin, the pole $p$ and the point $q$ with the sphere. By undoing the symmetry steps, we see the plane rotates along with the geodesic, and so we could have constructed the geodesic with the plane intersection in it's original place.

Similar arguments apply for the surfaces embedded in Minkowski, with the caveat that they are invariant under actions of the Lorentz group of the immersion space (which include spatial and hyperbolic rotations depending on the axis), and that for Lorentzian surfaces we need to disambiguate by type of geodesic.

\end{appendices}

\bibliography{bibliography}%
\end{document}